# Effect of Na(I) on Bioavailability for Cr(VI)_ and Cr(III)_ Arthrobacter species

E.Gelagutashvili, O.Rcheulishvili, A.Rcheulishvili

*Iv. Javakhishvili Tbilisi State University*
*E. Andronikashvili Institute of Physics*
*0177, 6, Tamarashvili St.,*
*Tbilisi, Georgia*

## Abstract

The biosorption of Cr(VI)_ and Cr(III)_ of Arthrobacter species (*Arthrobacter globiformis* and *Arthrobacter oxidas*) was studied application dialysis and atomic absorption analysis at various Na(I) concentrations. Is was shown significant difference between the binding constants as for Cr(VI)_ and Cr(III)_*Arthrobacter oxidas*, as well as Cr(VI)_ and Cr(III)_ *Arthrobacter globiformis* at various Na(I) concentrations.
It was shown, that bioavailability increases in both cases with decreases Na(I) concentration.

Key words: *Arthrobacter oxidas, Arthrobacter globiformis*, metal ions

## Introduction

Among the different heavy metals, chromium is one of the most toxic pollutants. This metal is introduced into natural waters through various industrial activities [1]. The two typical oxidative states of chromium in the environment are hexavalent, Cr(VI), and trivalent, Cr(III). Biotransformation of highly toxic and mutagenic hexavalent chromium [2] to relatively nontoxic trivalent Cr(III) form by chromate reducing bacteria offers an economical as well as ecofriendly option for chromium bioremediation.These two oxidation states have widely contrasting toxicity and transport characteristics: hexavalent chromium is more toxic, with high water solubility and mobility, while trivalent chromium is less soluble in water, less mobile and less harmful [3]. Due to the repulsive electrostatic interactions, Cr(VI) anion species are generally poorly adsorbed by the negatively charged soil particles and can move freely in the aqueous environments. In contrast, Cr(III) species normally carry positive electric charges and therefore can be easily adsorbed on the negatively charged soil particles [4].

Four chromate reducing bacterial strains, namely, *Arthrobacter sp. SUK*, *Arthrobacter sp. SUK 1205, Pseudomonas putida SKPD 1202*, and *Corynebacterium paurometabolum SKPD 1204* were previously isolated and reported from chromite mine overburden and mine seepage samples and found to reduce chromate during growth under aerobic conditions [5]. The process of chromate reduction is adversely affected by the presence of additional metal ions possibly due to metal toxicity and inhibition of the Cr(VI) reduction process [6].

Chromate reduction by viable cells of *Arthrobacter sp. SUK 1201* [7] was in general negatively affected when the reduction medium was supplemented with different heavy metals such as Ni(II), Zn(II), Mn(II), and Co(II) at equimolecular concentration. As compared to control, presence of Ni(II), Zn(II), Mn(II), and Co(II) showed nearly 66%, 74%, 60%, and 64% reduction, respectively [7]. However, Cr(VI) reducing capability of the isolate was enhanced when Cu(II) was present in the medium along with Cr(VI). Such stimulatory effect of Cu(II) on Cr(VI) reduction activity has also been reported for Cr(VI)-reduction by *Arthrobacter sp. SUK 1205* [8]. Many methods have been exploited to remove Cr (VI) in the environment, including empowering bacteria as bioremediation agent.

Effect of Na(I) on absorption Cr(VI)_ and Cr(III)_ of Arthrobacter species was studied in this paper using dialysis and atomic absorption analysis.

## Materials and Methods

The other reagents were used: NaCl, $CrCl_3$, $K_2CrO_4$. (Analytical grade). *Arthrobacter* bacterials were cultivated in the nutrient medium. Arthrobacter species cells were centrifuged at 12000 rpm for 10 min and washed three times with phosphate buffer (pH 7.0). The centrifuged cells were dried without the supernatant solution until constant weight. After solidification ( dehydrated) of cells (dry weight) solutions for dialysis were prepared by dissolving in phosphate buffer. This buffer was used in all experiments. A known quantity of dried bacterium suspension was contacted with solution containing a known concentration of metal ion. For biosorption isotherm studies, the dry cell weight was kept constant (1 mg/ml), while the initial chromium concentration in each sample was varied in the interval ($10^{-3}$ -$10^{-6}$ M). All experiments were carried out at ambient temperature. Metal was separated from the biomass with the membrane, which thickeness was 30μm Visking (serva) and analyzed by an atomic absorption spectrophotometer ,,Analyst-900'' (Perkin Elmer) $\lambda_{Cr}$=357.9 nm wavelength. Dialysis carried out during 72 h. Concentration of Na(I) 2mM, 20 mM , 50mM. The isotherm data were characterized by the Freundlich [9] equation, which by us in analogue cases were discussed in work [10].

## Results and Discussions

Biosorption of Cr ion in anion and cation forms for two kinds of *Arthrobacter* (*Arthrobacter globiformis 151B and Arthrobacter oxidas 61B*) at room temperature at various Na(I) concentrations were studied. Freundlich parameters evaluated from the isotherms with the correlation coefficients are given in table 1. As seen from table 1, the change in sodium concentration strongly affects as Cr (III) _*Arthrobacter* species and also Cr(VI)-*Arthrobacter* species complexes. In particular, as the concentration of sodium increases, the binding constant decreases in all cases. Comparative biosorption characteristics for Cr(III) *Arthrobacter* species shown (table 1) , that more decrease in bioavailability has been observed experimentally for Cr(III)-*Arthrobacter oxidas* as compared with *Arthrobacter globiformis*. This change is more pronounced for Cr (VI) _*Arthrobacter globiformis 151B* compared to Cr (VI) _*Arthrobacter oxidas 61B*. (Decreases from 3.8 x10-4 to 2.51 x10-4 in the case of Cr (VI) _*Arthrobacter oxidas 61B,* and for Cr (VI) -*Arthrobacter globiformis* 151B 2.09 x10-4 to 0.95 x10-4). Our results indicated that Cr(VI) and Cr(III) sorption at various Na(I) concentrations is depended of species of bacterial *Arthrobacter*. Differences between *Arthrobacter* species in metal ion binding may be due to the properties of the metal sorbates and the properties of bacterium (functional groups, structure and surface area, depending on the species). Functional groups within the wall provide the amino, carboxylic, sulfydryl, phosphate, and thiol groups that can bind metals [11]**.** It was shown, that the carboxyl groups were the main binding site in the cell wall of gram positive bacteria [12]. Therefore, it can be concluded that sodium ions "screen" chlorine ions to contact the active centers of the bacterium, in particular carboxyl groups. Gram-positive bacteria have also a greater sorptive capacity due to their thicker layer of peptidoglycan which contains numerous sorptive sites [13]. In our case  $n$  values  which reflects the intensity of sorption presents the same trend  for Cr(III)_ and Cr(VI)- *Arthrobacter  globiformis 151B* (from 1.69 to 1.59, from 1.49 to 1.05 respectively   with increase Na(I) concentration)  but, as seen from table 1 for Cr(III)_ and Cr(VI)_ *Arthrobacter oxidas*  received sorption intensity indicator are  different.

Table 1. Biosorption parameters for Cr(III)_ and Cr(VI)_*Arthrobacter* species at various Na$^+$ concentration

| | [Na$^+$], mM | Biosorption constant, K×10$^{-4}$ | Absorption capacity, n | Correlation coefficient R$^2$ |
|---|---|---|---|---|
| Cr(III)-*Arthrobacter oxidas 61B* | 50 | 2.1 | 2.17 | 0.98 |
| | 20 | 4.1 | 4.34 | 0.92 |
| | 2 | 8.7 | 1.56 | 0.97 |
| | without Na$^+$ [10] | 26.0 | 1.37 | 0.98 |
| Cr(III)-*Arthrobacter globiformis 151B* | 50 | 3.2 | 1.59 | 0.98 |
| | 20 | 4.6 | 1.75 | 0.94 |
| | 2 | 8.0 | 1.69 | 0.96 |
| | without Na$^+$ [10] | 20.2 | 1.23 | 0.98 |
| Cr(VI) *Arthrobacter oxidas 61B* | 50 | 2.51 | 1.92 | 0.98 |
| | 20 | 3.23 | 1.58 | 0.99 |
| | 2 | 3.8 | 1.03 | 0.93 |
| | without Na$^+$ [10] | 4.6 | 1.25 | 0.98 |
| Cr(VI)-*Arthrobacter globiformis 151B* | 50 | 0.95 | 1.05 | 0.97 |
| | 20 | 1.59 | 1.2 | 0.95 |
| | 2 | 2.09 | 1.49 | 0.99 |
| | without Na$^+$ [10] | 3.4 | 1.35 | 0.96 |

The effect of  *pH* on Cr(VI) reduction and removal from   aqueous solution was studied in the range of 1-4. *Arthrobacter viscosus* biomass was used for Cr(VI)  biosorption[14]. The best removal efficiency and uptake were  reached at *pH 4* [14]. *Arthrobacter* sp. SUK 1201, a potent isolate reported from chromite mine overburden of Orissa, India, has been evaluated for Cr(VI) reduction with immobilized whole cells. Optimum *pH* for Cr(VI) reduction was 7.0, and the process was inhibited by metal ions such as Ni(II), Co(II), Cd(II), Zn(II), and Mn(II) but not by Cu(II) and Fe(III) [7].

Binding constants for Cr(III)-*Arthrobacter* species in all cases are more than for Cr(VI)_*Arhtobacter* species. Similar results were obtained without Na (I) [10], but the difference was more significant than in the presence of Na (I). In particular, without Na (I), the biosorption constant for Cr (III) _*Arthrobacter oxidas*  and Cr (III) _*Arthrobacter globiformis*  is approximately 6 times greater than for Cr (VI) _*Arthrobacter oxidas*  and Cr (VI) _*Arthrobacter globiformis*. At large concentrations of Na (I) ions (50 mmol) and without it, the difference between the binding constants for Cr (III)_*Arthrobacter oxydas* is 12 times greater and for Cr (III) _ *Arthrobacter globiformis*  6-fold greater.  In a same case for Cr (VI) _*Arthrobacter oxydas* the difference is only 2-fold greater and for Cr (VI) -*Arthrobacter globiformis* 4-fold greater.

Thus, effect of Na(I) ions are more significant in the case Cr(III)_*Arthrobacter* species ,than for Cr(VI)_*Arthrobacter* species. All this is natural considering that Cr(III) positively charged ions as Na(I) and both have the same binding active centers in contrast to negatively charged Cr(VI) ions.